\documentclass[11pt]{article} 

\usepackage[utf8]{inputenc} 

\usepackage{geometry} 
\usepackage{graphicx} 

\usepackage{booktabs} 
\usepackage{array} 
\usepackage{paralist} 
\usepackage{verbatim} 
\usepackage{subfig} 

\usepackage{fancyhdr} 
\usepackage{sectsty}
\allsectionsfont{\sffamily\mdseries\upshape} 

\usepackage[nottoc,notlof,notlot]{tocbibind} 
\usepackage[titles,subfigure]{tocloft} 

\title{A Mathematical Method for Deriving the Relative Effect of Serviceability on Default Risk}
\author{Graham Andersen\footnote{+61 438 696 600; graham.andersen@morgij.com.au}, David Chisholm\footnote{+61 438 562 359; david.chisholm@morgij.com.au\newline \newline \indent Morgij Pty Ltd \newline \indent L9, 146 Arthur St \newline \indent North Sydney Australia 2060}}

\date{23 November 2011} 

\begin{document}
\maketitle
\newpage
\section*{Abstract}
The writers propose a mathematical Method for deriving risk weights which describe how a borrower’s income, relative to their debt service obligations (serviceability) affects the probability of default of the loan.

The Method considers the borrower’s income not simply as a known quantity at the time the loan is made, but as an uncertain quantity following a statistical distribution at some later point in the life of the loan.  This allows a probability to be associated with an income level leading to default, so that the relative risk associated with different serviceability levels can be quantified.  In a sense, the Method can be thought of as an extension of the Merton Model to quantities that fail to satisfy Merton’s ‘critical’ assumptions relating to the efficient markets hypothesis.

A set of numerical examples of risk weights derived using the Method suggest that serviceability may be under-represented as a risk factor in many mortgage credit risk models.
\newpage
\section{Introduction}
Traditional mortgage credit risk models take a functional approach to producing a credit risk metric, generating an absolute final risk metric from the starting point of absolute base probability of default and loss given default. The authors have instead developed a differential credit risk model which is based on how the risk metric relates to changing loan characteristics, leaving the decision about absolute risk levels to the user.

The authors propose a mathematical Method for deriving risk weights which describe how a borrower’s income, relative to debt service obligations (serviceability) affects the probability of default of a loan.  The Method considers the borrower’s income not simply as a known (assessed) quantity at the time the loan is made, but as an uncertain quantity following a statistical distribution at some later point in the life of the loan.  This follows a similar path to the Merton model \cite{mer}, often used in the analysis of corporate credit risk.  Treating income as a distribution allows a probability to be associated with an income level leading to default, so that the relative risk associated with different serviceability levels can be quantified. We provide numerical examples of serviceability risk weights, and suggest that current industry approaches may significantly under-represent the importance of serviceability to mortgage credit risk.

The Method can be extended so that other risk characteristics of the borrower can be treated as modifications to the income distribution chosen, rather than treated as discrete modules within the credit risk model.

The Method was developed in the context of residential mortgages, but has applicability to any form of lending supported by an income stream which is subject to uncertainty.

Although not addressed in detail in this paper, the Method can also be applied to other loan characteristics where a variability over time can lead to default (such as loan-to-value ratio in a limited recourse environment) to expand the applications for differential credit risk models. 
\newpage
\section{Background}

\subsection{Credit Risk Models}

The traditional structure of most mortgage credit risk models can be thought of as a process whereby a metric of credit risk is derived for a specified borrower via the following steps:
\begin{enumerate}
\item A base probability of default (“Base PD”) and base loss given default (“Base LGD”) are defined, corresponding to some generic borrower, and the credit risk appetite and view of the lender;
\item The Base PD and Base LGD are adjusted to reflect various characteristics of the specified loan, such as borrower or collateral quality (this resulting in “Adjusted PD” and “Adjusted LGD”).  Adjustments are typically applied as successive multiplicative factors (“Risk Weights”), each corresponding to a particular risk characteristic of the borrower\footnote{Typically, the Base PD will be further adjusted to fall within certain maximum and minimum levels, however this process is not relevant to this paper};
\item The product of Adjusted PD and Adjusted LGD give an expected loss (“EL”)\footnote{Some models start with a Base EL in step 1, adjusted as per step 2 and omit this step 3.  This is clearly mathematically equivalent to the approach described.};
\item EL can be used to drive the lending decision, pricing and capital requirement.
\end{enumerate}

Structured finance RMBS criteria of the major credit rating agencies are good public examples of this approach.

It is important to note in relation to point 2, that each Risk Weight is typically determined by a single discrete risk characteristic (or small closely related subset of characteristics).  This ‘Modularity Assumption’ greatly simplifies the mathematical structure of a credit risk model, and is generally found to still lead to sensible results.

Within this traditional framework, many variations exist, however the basic structure is close to universal.

\subsection{The Merton Model}
The analysis of corporate credit risk is often framed in terms of the Merton Model \cite{mer}, where the value of a security issued by a firm is given in terms of the value of the firm, with that value assumed to be subject to continuous volatility.  The essence of the Merton Model is that default is taken to correspond to a deficiency of assets.

The characteristics of mortgages do not naturally lend themselves to a Merton-style analysis, since:
\begin {enumerate}
\item The underlying asset (the security property) is not a fungible asset traded in an open market with continuous price discovery; and
\item Mortgages rely primarily on the income of the borrower and only secondarily on the security property value.
\end {enumerate}

This paper focuses on the second issue, the relationship between the income of the borrower and probability of default.

Sy \cite{sy} applied the Merton Model to borrower income to derive probabilities of default, as part of a programme of developing a credit model integrating probability of default and loss given default, however the authors are not aware of any subsequent developments along these lines in the literature.

\subsection{Absolute vs Differental Risk Models}
\subsubsection{Absolute Risk Models}
A hallmark of most mortgage credit risk models is that they are based on absolute measures of risk.  Base PD, Base LGD and each of the Risk Weights are given as fixed numerical parameters, and EL is simply a function of these.  While the parameters may be adjusted from time to time, at any time they each have a single value deemed ‘correct’ under the model\footnote{In the case of Credit Rating Agencies, the different rating bands can be thought of as multiple models, from which users may select the one reflecting their risk appetite}.  It follows that for each unique borrower defined by a particular set of risk characteristics, there is a single EL value, again deemed ‘correct’ under the model.

We suggest that this absolute approach to risk suffers a number of shortcomings:
\begin{itemize}
\item Much more attention is paid to the end points (Base PD, Base LGD and EL) than to the way that risk changes as a result of risk characteristics (Risk Weights and the underlying algorithm);
\item There is a tendency to validate the model by simply testing EL against historical loss experience rather than to develop a coherent rationale for each Risk Weight; and
\item Each model binds its user to a very specific risk view.
\end{itemize}

\subsubsection{A Differential Risk Model}
The writers have developed a proprietary credit risk model, initially for application to residential mortgages, the Risk Quantification Methodology (“RQM”).  The RQM is not an absolute model, but a differential model.  RQM does not specify any absolute base risk metric, only how the risk metric changes as loan characteristics change.  Individual users may specify their own base levels (and other parameters) to generate a model reflecting their own risk view.

In the course of developing RQM for mortgages, we have developed a general mathematical approach to deriving Risk Weights to reflect the borrower’s income level relative to their debt service obligations (serviceability). 

\section{Serviceability}
\subsection{Defining Serviceability}
We define serviceability as the ability of the borrower to meet their debt service obligations out of their net cash income\footnote{This is a deliberately general definition, reflecting the wide range of types of lending and borrower.  Our Method can be applied to whatever specific measure of serviceability applies to the particular type of lending and/or borrower}.

For the purposes of demonstrating our Method, we consider one common measure of serviceability in mortgage lending, the Net Servicing Ratio (“NSR”), defined as:
\begin{equation}
N=I_N/R_N
\end{equation}

\indent where
\begin{quote}
$N$ is the NSR,
\newline $I_N$ is the borrower’s assessed periodic net income at the time of application, incorporating whatever stresses are mandated by the lender, and
\newline $R_N$is the borrower’s minimum required repayment at the time of application, incorporating whatever stresses are mandated by the lender.
\end{quote}

Different lenders approach NSR differently, however typically lenders will apply various stresses to ensure that borrowers have some ‘cushion’ against changing circumstances, such as interest rate rises or income reduction.  Typical stresses include:
\begin{itemize}
\item Only crediting a proportion of the borrowers actual net income to $I_N$.  We will refer to the actual (unstressed) assessed income as $I_0$;
\item Applying an interest rate stress to calculate the minimum required repayment, $R_N$;
\item Requiring all borrowers to meet a particular NSR threshold (eg, 1.1).
\end{itemize}

\subsection{Defining Repayment Coverage and Default}
While NSR measures serviceability at the point of loan application and approval, we also need an ongoing measure of serviceability from time to time.  We define the Repayment Coverage Ratio (“RCR”) as:
\begin{equation}
C=I/R
\end{equation}

\indent where
\begin{quote}
C is the RCR at a given time,
\newline I is the borrower’s actual periodic net income at that time, and
\newline R is the borrower’s actual required repayment at that time.
\end{quote}

Now we can assume that the loan will default if the RCR is less than 1, meaning that the borrower’s net income is insufficient to meet their repayment obligation.

\subsection{Changes in Net Income and Repayment Coverage Ratio}
While a borrower’s actual periodic net income $I$ may be reasonably accurately assessed at the time of application, it can change over the life of the loan (either increasing or decreasing), for a variety of reasons.  Focussing on individual borrowers these could include:
\begin{itemize}
\item Pay rises or reductions, say through promotion, demotion or career change
\item Family income reduction with maternity or paternity leave;
\item Increased or decreased expenditure through lifestyle change, say from child-raising or ‘sea-change’;
\item Unforeseen life events such as illness, divorce or business success.
\end{itemize}

\subsection{Income as a Distribution}
We are interested in default behaviour over the life of the loan, so we need to consider the borrower’s net income over the life of the loan.  We achieve this by considering the borrower’s ‘true’ income over time not as a fixed quantity, but as variable quantity, ‘migrating’ from its initial value.  This leads us to consider it as a distribution around I0, the initial, unstressed, assessed income.

We need to clarify the timeframe over which we are considering the income migration. This may appear to be the legal tenor of the loan, however this will not always be the case.  For example, residential mortgages typically have tenors of around 30 years.  However default behaviour is heavily concentrated in the first few years.  The Method restricts the relevant timeframe, the Default Horizon, which depends on the particular features of the loan and may be less than its legal tenor.

To summarise, under the Method, a borrower’s income distribution reflects the likelihood of the borrower’s income migrating to another value within the Default Horizon.

For the purposes of demonstrating the Method below, we make the assumption that true income over time is distributed normally around the assessed income, but note that the Method can be used with other choices of distribution, including non-central distributions (though with some loss of mathematical simplicity).

\subsubsection{Comparison to Merton and Sy}
We note that our approach differs from Sy \cite{sy} and Merton \cite{mer} in that we do not model the income distribution as the result of a continuous diffusion process, but instead treat it simply as a defined distribution within the default horizon.  This is equivalent to discarding Merton’s assumptions that:
\begin{enumerate}
\item The underlying quantity is a continuously traded asset; and
\item The price of the underlying quantity can be described by a particular diffusion-type stochastic process consistent with the efficient markets hypothesis.
\end{enumerate}

Clearly a person’s income is generally not continuously traded in any market in any meaningful sense, let alone an efficient one conforming to Merton’s requirements.  Accordingly, neither of these assumptions can hold in relation to the analysis of borrower incomes.

Merton describes both of these assumptions as critical, and the detailed mathematical framework of the Merton Model does indeed depend on them.  Nevertheless, we show below that a meaningful analysis can still be developed from our weakened starting point.

\section{Risk Weights for Serviceability}
We can now derive Risk Weights for serviceability, based on the treatment of net income as a distribution.

\subsection{Key Assumptions}
\subsubsection{Income Stress}
In line with common lender practice, we assume that the initial unstressed assessed income $I_0$ is related to the initial stressed income $I_N$ by some stress factor, $f\leq 1$.
\begin{equation}
I_N=fI_0
\end{equation}

\subsection{Income Distribution}
We assume that ‘true’ net income, $I$,  within the Default Horizon, is a normally distributed quantity around $I_0$, with standard deviation equal to some fixed percentage $s$ of $I_0$.
We represent this as $I(p)$, where p is the probability of the actual income within the Default Horizon being at least $I(p)$.  For example, $I(0.5)$ is I, and $I(0.7)$ is the income of the 30’th centile borrower.

\begin{equation}
I(p)=InvN(p,sI_0,I_0))
\end{equation}

\indent or

\begin{equation}
p(I)=Norm(I,sI_0,I_0))
\end{equation}

\indent where
\begin{quote}
$InvN$ is the inverse normal distribution function with arguments (probability, standard deviation, mean) and,
\newline $Norm$ is the cumulative normal distribution function with arguments (value, standard deviation, mean).
\end{quote}

\subsubsection{Repayment Obligation}
$R$ (the borrower’s actual required repayment) can be subject to change, for example as a result of interest rate changes.  Rather than consider $R$ as a distribution, we will assume that $R = R_N$, that is, the lender’s repayment stress assumption is reached.  This means that the Risk Weights derived relate to the NSR-related risk under the lender’s stressed repayment assumptions.

\subsection{Derivation of Risk Weights}
Since income is a distribution, our repayment coverage, $C$, also becomes a distribution.  (Note that NSR does not, as it is defined in terms of the assessed $I_0$ and $R_N$.)

\begin{equation}
C(p)=I(p)/R
\end{equation}

There is a probability, $p_D$, that the ‘true’ income level within the Default Horizon is such that our specified repayment stress (defining $R_N$) at that time leaves the borrower with a repayment coverage of exactly 1.  We can consider $p_D$ to be a measure of the NSR-dependent probability of default for that borrower, under that repayment stress.  $p_D$ is defined by:

\begin{equation}
C(p_D)=I(p_D)/R_N\equiv 1
\end{equation}

\indent or

\begin{equation}
I(p_D)=R_N=fI_0/N
\end{equation}

We can then express $p_D$ for a given NSR directly as

\begin{equation}
p_D(N)=Norm(fI_0/N,sI_0,I_0)
\end{equation}

The Risk Weight for a given NSR can now be expressed as the ratio between the $p_D$ for that NSR and $p_D$ for some base NSR.  We choose 1.0 as our base NSR:

\begin{equation}
F_N=\frac{p_D(N)}{p_D(N=1)}
\end{equation}

\indent or

\begin{equation}
F_N=\frac{Norm(fI_0/N,sI_0,I_0)}{Norm(fI_0,sI_0,I_0)} \label{eq:E1}
\end{equation}

Now, the form of the normal distribution function is

\begin{equation}
Norm(aI_0,sI_0,I_0)=c\int^{aI_0}_{-\infty}exp\left[\frac{-(I-I_0)^2}{2(sI_0)^2}\right]dI
\end{equation}

where c is a constant.

Dividing through the exponential by $I_0^2$ and substituting $I'=I/I_0$ gives

\begin{equation}
Norm(aI_0,sI_0,I_0)=c\int^{a}_{-\infty}exp\left[\frac{-(I'-1)^2}{2s^2}\right]dI'=Norm(a,s,1)
\end{equation}

which shows that equation \ref{eq:E1} can be simplified to
\begin{equation}
F_N=\frac{Norm(f/N,s,1)}{Norm(f,s,1)} \label{eq:E2}
\end{equation}

At this point, we have derived an expression for ‘theoretical’ factors\footnote{We would expect that theoretical factors might be adjusted for various practical reasons in setting actual factors for a working credit risk model.} $F_N$ based only on:
\begin{itemize}
\item The income stress factor,$f$;
\item NSR, $N$; and
\item The standard deviation of ‘true’ income, expressed as a percentage of the ‘true’ income, $s$.
\end{itemize}

In particular, the Risk Weights do \textit{not} explicitly depend on:
\begin{itemize}
\item The borrower rate at the time NSR is calculated;
\item The interest rate stress in the NSR calculation;
\item The required repayment; or
\item The borrower's reported income\footnote{As mentioned above, the Method can be used with income distributions other than the normal distribution.  We note that a formulation of factors independent of these elements may not always be possible.}.
\end{itemize}

\subsection{The Meaning of $p_D$}
We characterise $p_D$ above as “a measure of the NSR-dependent probability of default”.  Does it make sense to then derive $p_D$ directly for desired cases, rather than just the Risk Weights $F_N$?  We caution that direct calculations of $p_D$ using the Method are not necessarily meaningful.

Consider more carefully what $p_D$ represents.  It is a measure of the probability of default, based solely on NSR and \textit{ignoring the values of all other characteristics of the mortgage}.  Specifically, $p_D$ is the same for any two mortgages with the same NSR, regardless of differences in such fundamental characteristics as loan-to-value ratio, or borrower credit history.  Under the Method we are calculating only the ratios that define Risk Weights (since the effects of these other characteristics cancel out) and not any observable probability of default.  Note that we are implicitly using the Modularity Assumption referred to above.  Specifically, we have assumed that the Risk Weights for NSR depend only on NSR and not on any other risk characteristics.

Note that the approach of treating the Risk Weights as the derived quantities of interest, is fundamentally different from Sy \cite{sy}, who treated probabilities of default as the quantities of interest.

\subsection{Income Distributions and Other Risk Characteristics}
The Method as described here considers serviceability as a discrete risk characteristic, and income as having a single given distribution.

We observe that many borrower characteristics typically used in mortgage credit risk models could be treated under our Method as secondary parameters affecting the income distribution, rather than primary factors directly affecting probability of default.  We suggest that this could result in a more internally coherent credit risk model, while accepting that the construction of such a model would appear to require the development of a detailed theory of borrower income behaviour.

Some possible examples are:

\begin{itemize}
\item A borrower with history of defaults or bankruptcy might be associated with a negatively skewed distribution (ie, their income is more likely to fall than rise);
\item A self-employed borrower might be associated with a wider distribution than a full-time borrower (ie, they are more exposed to both upside and downside in income);
\item A young professional borrower might be associated with a positively skewed income (ie, their income is more likely to rise than fall).
\end{itemize}

\section{Numerical Examples}
We now provide some specific numerical examples of NSR Risk Weights.

Our examples are based on an income stress, $f$, of 90\%.  Table 1 below shows NSR-related Risk Weights tabulated by NSR, N and income standard deviation, $s$, calculated according to equation \ref{eq:E2}.

We particularly note two features of the numerical examples.

Firstly, the range of Risk Weights is very wide, over orders of magnitude, suggesting serviceability is a powerful risk determinant.

Secondly, small changes of NSR around 1.0 result in significant changes in risk.  For example, at 30\% standard deviation, an increase in NSR from 1.0 to 1.1 reduces the probability of default by 26\%.

\begin{table}
\centering
\caption{Numerical Examples of Risk Weights}
\begin{tabular}{c| c c c c c c c}
\hline \hline
\multicolumn{1}{c}{}& \multicolumn{7}{l}{\textbf{Income SD}}\\
\hline
\bfseries NSR & \bfseries 10\% & \bfseries 15\% & \bfseries 20\% & \bfseries 25\% & \bfseries 30\% & \bfseries 35\% & \bfseries 40\% \\
\hline
\bfseries 0.2 & 6.30 & 3.96 & 3.24 & 2.90 & 2.71 & 2.58 & 2.49 \\
\bfseries 0.3 & 6.30 & 3.96 & 3.24 & 2.90 & 2.71 & 2.58 & 2.49\\
\bfseries 0.4 & 6.30 & 3.96 & 3.24 & 2.90 & 2.71 & 2.58 & 2.49\\
\bfseries 0.5 & 6.30 & 3.96 & 3.24 & 2.90 & 2.70 & 2.55 & 2.44\\
\bfseries 0.6 & 6.30 & 3.96 & 3.22 & 2.84 & 2.58 & 2.38 & 2.23\\
\bfseries 0.7 & 6.29 & 3.85 & 2.99 & 2.53 & 2.25 & 2.05 & 1.90\\
\bfseries 0.8 & 5.64 & 3.16 & 2.38 & 2.01 & 1.79 & 1.65 & 1.55\\
\bfseries 0.9 & 3.15 & 1.98 & 1.62 & 1.45 & 1.35 & 1.29 & 1.25\\
\bfseries 1.0 & 1.00 & 1.00 & 1.00 & 1.00 & 1.00 & 1.00 & 1.00\\
\bfseries 1.1 & 0.22 & 0.45 & 0.59 & 0.68 & 0.74 & 0.78 & 0.81\\
\bfseries 1.2 & 0.04 & 0.19 & 0.34 & 0.46 & 0.55 & 0.61 & 0.66\\
\bfseries 1.3 & 0.01 & 0.08 & 0.20 & 0.32 & 0.41 & 0.49 & 0.55\\
\bfseries 1.4 & 0.00 & 0.03 & 0.12 & 0.22 & 0.32 & 0.40 & 0.46\\
\bfseries 1.5 & 0.00 & 0.02 & 0.07 & 0.16 & 0.25 & 0.33 & 0.40\\
\bfseries 1.6 & 0.00 & 0.01 & 0.05 & 0.12 & 0.20 & 0.27 & 0.34\\
\bfseries 1.7 & 0.00 & 0.00 & 0.03 & 0.09 & 0.16 & 0.23 & 0.30\\
\bfseries 1.8 & 0.00 & 0.00 & 0.02 & 0.07 & 0.13 & 0.20 & 0.26\\
\bfseries 1.9 & 0.00 & 0.00 & 0.01 & 0.05 & 0.11 & 0.17 & 0.23\\
\bfseries 2.0 & 0.00 & 0.00 & 0.01 & 0.04 & 0.09 & 0.15 & 0.21\\
\hline
\end{tabular}

\end{table}

Readers may object to the inclusion of NSR less than 1 on the grounds that such loans would not be made.  We have included these firstly for completeness, but also because different lenders can define NSR with different repayment assumptions and stresses.  Accordingly it is possible for one lender to assign a borrower an NSR less than 1, while another assigns an NSR of 1, and the borrower has an initial repayment coverage greater than 1.  For example a borrower may be assessed and approved on a greatly reduced initial honeymoon repayment, while the ‘true’ NSR should more properly be assessed on reversion to normal repayment terms.

\section{Existing Approaches}
Approaches to serviceability are hard to find publicly documented.  However, anecdotally lenders tend to approach serviceability on a threshold basis for all borrower types.  That is, a borrower must satisfy some minimum requirement (which may vary between different types of borrower), but no or limited credit is given for excess income.

This appears at odds with the results in the Method outlined above, and the numerical examples. If serviceability is so important, should lenders have noticed these results and started using more differentiated risk weights?

We suggest some potential explanations.

\subsection{Rising Asset Values}
In a benign economic environment (often regarded as 'normal') where credit is cheaply available and the value of assets used as security are rising, then much default behaviour can be ‘masked’ by the successful liquidation of the security, largely eliminating the need to worry about serviceability beyond minimum threshold requirements.  Of course, once asset prices start falling and credit is rationed, the risk mitigation of overcollaterisation disappears and serviceability becomes key.

\subsection{Threshold Incentives}
When a lender imposes a serviceability threshold, this can create an incentive to lend up to that threshold.  In simple terms, lender revenues are maximised by lending to the threshold, while borrower utility (if thought of as the ability to obtain credit and purchase assets) is similarly maximised.  A simple serviceability threshold discourages deeper examinations of the risks associated with serviceability, or even the collection and storage of data.

\section{Other Applications}
While this paper has focused on analysing serviceability, we suggest that the mathematical approach described is not necessarily restricted to this risk characteristic.  In fact, the same approach could be applied to any quantity which:
\begin{itemize}
\item Is known at the time of application and settlement;
\item Subsequently varies unpredictably, but can be meaningfully thought of as following a distribution;
\item Induces a default once it reaches a certain value; and
\item Does not satisfy the critical assumptions of the Merton Model.
\end{itemize}

An example might be loan-to-value ratio in a limited recourse environment, where a borrower is incentivised to default once the perceived property value falls below a threshold percentage of the outstanding debt.  While at first sight this might seem to simply restate Merton’s original proposition in the context of mortgages, it differs in two respects:
\begin{enumerate}
\item The market for residential property does not appear to fulfil the critical assumptions of the Merton Model, leading to our different treatment; and
\item The default is not treated as a simple asset insufficiency, but as driven by borrower behavior, potentially leading to quite different numerical assumptions and outcomes.
\end{enumerate}

\section{Regulatory Issues}
Of late, regulators are increasingly focused on how to ensure that regulatory capital is appropriately set ‘through the cycle’.  To this end, the Basel III rules include measures intended to

\begin{quote}
“dampen any excess cyclicality of the minimum capital requirement;

promote more forward looking provisions;

conserve capital to build buffers at individual banks and the banking sector that can be used in stress; and

achieve the broader macroprudential goal of protecting the banking sector from periods of excess credit growth" \cite[para19]{bas}
\end{quote}

A key difficulty of implementing such a framework is that, as observed above, rising asset values and benign economic conditions can mask default behavior, so that historical data alone is not an appropriate guide to setting capital requirements.

We suggest that introducing a capital allocation or stress based on serviceability, such as could be based on the Method, would provide a natural counter-cyclical  component to capital requirements.  Simply, while rising asset values and benign economic conditions tend to encourage reductions in capital based on low recent losses, a serviceability weight to capital would naturally increase as borrower gearing increases \textit{regardless of the value of the asset securing the loan}.

\section{Acknowledgement}
The authors acknowledge helpful comments and insights from David Howard-Jones of Oliver Wyman.


\begin{thebibliography}{9}
\bibitem{mer} Merton, R.C. (1974). On the Pricing of Corporate Debt: The Risk Structure of Interest Rates. \textit{Journal of Finance, 29 (2)}, 449-470.
\bibitem{sy} Sy, W. (2007, October). textit{Research paper 204: A Causal Framework for Credit Default Theory.} Available at the Australian Prudential Regulatory Authority web site: http://www.apra.gov.au/Policy/upload/A-casual-framework-for-credit-default-theory-September-2007.pdf, last accessed on 19 May 2011.
\bibitem{bas} Basel Committee on Banking Supervision (2010, December, revised 2011, June). \textit{Basel III: A global regulatory framework for more resilient banks and banking systems.} Bank for International Settlements.
\end{thebibliography}
\end{document}